Review of several false positive error rate estimates for latent fingerprint examination proposed based on the 2014 Miami Dade Police Department study

Madeline Ausdemore[1], Jessie H. Hendricks[1], Cedric Neumann[1]

[1] Department of Mathematics and Statistics, South Dakota State University, 57007 Brookings, SD

Contact:
Dr. Cedric Neumann
Associate Professor of Statistics
Department of Mathematics and Statistics
South Dakota State University
Box 2225
57007 Brookings, SD
Cedric.Neumann@me.com

This project was supported in part by Award No. 2014-IJ-CX-K088 awarded by the National Institute of Justice, Office of Justice Programs, U.S. Department of Justice. The opinions, findings, and conclusions or recommendations expressed in this paper are those of the authors and do not necessarily reflect those of the Department of Justice.

## I. Introduction

During the past decade, several studies have been conducted to estimate the false positive error rate (FPR) associated with latent fingerprint examination. The so-called "Black-box study" by Ulery et al. [1] is regularly used to support the claim that the FPR in fingerprint examination is reasonably low (0.1%). Ulery et al.'s estimate of the FPR is supported by the results of the extensive study of the overall fingerprint examination process by Langenburg [2].

In 2014, the Miami-Dade Police Department (MDPD) Forensic Services Bureau conducted research to study the false positive error rate associated with latent fingerprint examination [3]. They report that approximately 3.0% of latent fingerprint examinations result in a false positive conclusion. Their estimate of the FPR becomes as high as 4.2% when inconclusive decisions are excluded from the calculation. In their 2016 report, the President's Council of Advisors on Science and Technology (PCAST) proposes that the MDPD FPR estimate be used to inform jurors that

Page 1 of 25

errors occur at a detectable rate in fingerprint examination; more specifically, they declare that false positives may occur as often as 1 in 18 cases [4, pp. 94-97].

The large discrepancy between the FPR estimates reported by Ulery et al. [1] and Langenburg [2] on the one hand, and the MDPD on the other hand, causes a great deal of controversy. For example, a recent Canadian case study advocates for a re-analysis of the MDPD data and of its interpretation by PCAST, before the PCAST point of view "becomes an urban myth" [5]. The Organization of Scientific Area Committees Friction Ridge subcommittee (OSAC FRS) [6] proposes an alternative estimate of the MDPD FPR. In this paper, we review the MDPD study and the various error rate calculations that have been proposed to interpret its data. To assess the appropriateness of the different proposed estimates, we develop a model that re-creates the MDPD study. This model allows us to estimate the expected number of false positive conclusions that should be obtained with any proposed FPR and compare this number to the actual number of erroneous identifications observed by MDPD.

**II.     Review of Pacheco et al. [3]**

In 2014, the MDPD reported the results of a study designed to evaluate the accuracy and reliability of fingerprint examination conducted using the standard ACE-V procedure [3]. This study consisted in three phases. Phases 1 and 2 were designed to study the *Analysis, Comparison* and *Evaluation* stages of the ACE-V process, while phase 3 was designed to study the *Verification* stage [3, pp. 22-34].

The following paragraphs summarise the design and results obtained by Pacheco et al. [3]. The page numbers in the MDPD report have been added for convenience. The design is based on a set of 80 latent prints that were collected and partitioned into two disjoint subsets consisting in 40 prints each, where each subset included latent prints of varying difficulty [3, pp. 23-28]. In addition, a full set of control prints (consisting in impressions from all ten fingers and from both palms) was collected from ten individuals [3, pp. 23, 28]. Control prints from the true sources were available for 56 out of the 80 test latent prints, and absent for the remaining 24. The repartition of these 56/24 latent prints across the two subsets of 40 is unknown.

In phase 1 of the study, the participants were provided with identical packets containing the first group of 40 unknown latent prints and all 10 sets of control prints [3, pp. 29-31]. For each of the 40 latent prints, participants were instructed to first perform an initial analysis to determine if the latent print was suitable for comparison purposes (or, *of value*). If the latent print was deemed to be of value, the participants were instructed to compare the print to three pre-specified sets of control prints (from the ten sets available for the study) and make a conclusion of *identification*, *exclusion*, or *inconclusive* (control prints from the source of the latent print were not necessarily present in these three specified sets of prints). In phase 2, the 109 participants who returned their phase 1 packets were divided into two groups, A and B [3, p. 31]. The second set of 40 latent prints (that had not been used in phase 1) was further divided into two sets of 20 latent prints each. Individuals placed in group A were provided with identical packets containing the first set of 20 latent prints and the 10 sets of control prints. Similarly, the individuals placed in group B were provided with identical packets containing the second set of 20 latent prints and the full set of control prints. The participants then followed the same process outlined for phase 1. Finally, in phase 3, participants were presented with sets of prints that consisted in the "identifications" made in phase 2 (regardless of whether or not the "identification" was true) [3, pp. 31-34]. The participants were then asked to verify these "identifications" by indicating if they agreed or disagreed with the conclusions made in phase 2. If the participant believed there was insufficient information to identify or exclude, they were instructed to make an inconclusive decision. To test for bias and repeatability, the packets constructed for this phase were not identical. A summary of the data collection methods is given on pages 36-39 of the MDPD report. We are not concerned with phase 3 of the MDPD study and it is not considered further in the current paper.

Phase 1 packets were sent to 140 participants. A total of 109 packets with results from phase 1 were returned [3, p. 38], including a total of 4233 decisions [3, p. 52, table 2] out of a possible 4360 decisions (109 packets × 40 latent prints per packet). This indicates that some participants in the study returned results for some of the latent prints provided in the packets and failed to return results for others. Out of the 4233 decisions that were returned, 1023 latent prints were deemed of no value. The remaining 3210 latent prints were compared to the pre-specified sets of control prints and a conclusion about their source was reached [3, p. 52, table 2].

Phase 2 packets were distributed to all 109 respondents from phase 1 [3, p. 38]. Results were returned by 88 participants in phase 2 [3, p. 38], including 1730 decisions [3, p. 52, table 2] out of a possible 1760 decisions (88 packets × 20 latent prints per packet). Out of the 1730 decisions that were returned from phase 2, 1342 latent prints were deemed of value for comparison and the remaining 388 were deemed of no value for comparison [3, p. 52, table 2].

The design of the MDPD study involves the comparison of each latent print (if it is deemed of value for comparison) to 3 sets of control prints. This design is closer to casework conditions where latent print examiners have to "search" for corresponding control impressions. This is different from the study by Ulery et al. [1], where each latent print was paired with a single control print. With the Ulery et al. design, only two types of error are possible during the evaluation stage of ACE-V: erroneous identifications and erroneous exclusions (assuming that inconclusive decisions are not considered to be 'errors'). With the MDPD design, a third type of error is possible: it is also possible to report the wrong finger of the correct person as the source of a latent print (when the control prints of the true donor were provided as part of the 3 sets of control prints).

### III. Error Rate estimates based on the MDPD data

In their report, the MDPD claims an FPR estimate of 3.0% [3, p. 53, table 4]. This rate was obtained by relating the 42 erroneous identifications observed during the study to a total of 1398 decisions (including inconclusive conclusions). The MDPD estimate raises to 4.2% if the inconclusive conclusions are not accounted for (42 erroneous identifications out of 1398 – 403 = 995 decisions) [3, p. 53, table 4].

The original data used in this calculation can be seen in table 1, which corresponds to a modified version of table 11 in the MDPD report [3, p. 56].

| Source Present (Y/N) | # of Latent Prints | # of Decisions | Correct ID | Erroneous ID | | Inc. | Correct Excl. | Erroneous Excl. |
| | | | | Correct person, wrong finger | Incorrect person | | | |
|---|---|---|---|---|---|---|---|---|
| Yes | 56 | 3177 | 2457 | 35 | 4 | 446 | N/A | 235 |
| No | 24 | 1359 | N/A | N/A | 3 | 403 | 953 | N/A |
| Totals | 80 | 4536 | 2457 | 35 | 7 | 849 | 953 | 235 |

**Table 1:** A modified version of table 11 in Miami-Dade report where the Erroneous ID column has been divided into erroneous identifications in which the correct individual was identified, but the incorrect finger was reported and erroneous identifications in which the wrong individual was identified.

MDPD reports that the 42 erroneous identifications consist of 35 cases in which an examiner correctly identifies the individual that produced the print but fails to associate the latent print with the correct finger [3, p. 64], and 7 cases in which an examiner incorrectly identifies the individual that produced the latent print (three of which occur when control impressions from the true source are not present in the pre-determined set of control prints) [3, p. 56].

MDPD considers these 42 erroneous identifications in the numerator of all of their calculations. Although these 42 cases consist of decisions that occur when control impressions from the true source are present in the comparison process and when they are not present in the comparison process, the denominator only considers the 1359 test cases for which these impressions are not present in the comparison process and the sole 39 decisions for which the control impressions were present but errors were made (1359 + 39 = 1398).

Thus, the main point of contention with the MDPD calculation is that, while most of the erroneous identifications are observed when control prints from the true source are present in the test packets, the FPR estimate does not account for the number of decisions that are rendered in this scenario. In other words, the scenario of the numerator does not match the scenario of the denominator of the FPR estimate. This issue is reinforced by the surprisingly large discrepancy between the proportion of erroneous identifications made when the source is present (39 erroneous ID out of 3,177 decisions or 1.2%) and when it is not (4 out of 1359 decisions or 0.3%). This discrepancy is noted by the authors of the study but not accounted for in their calculations and interpretation of the data [3, p. 64].

Despite these issues and that the MDPD study was never published in a peer-reviewed journal, the PCAST report relies heavily on these results to discuss the estimated error rate associated with latent fingerprint examination [4]. Furthermore, the PCAST report considers, using the upper bound of a 95% confidence interval on the FPR estimate, that the true FPR could be as high as 1 in 18 (or approximately 5.4%) cases [4, p. 95].

The OSAC FRS has raised the issue of the inconsistency between the numerator and the denominator of the FPR calculation in the MDPD study [6]. The OSAC FRS proposes that the calculation for the FPR considers all 42 erroneous identifications out of the total number of decisions made, excluding inconclusive decisions (4536 total decisions − 849 inconclusive decisions = 3,687) [6, p. 4]. This lowers the FPR to approximately 1.1%, with a 95% confidence interval upper bound of approximately 1.5%. The OSAC FRS does not expand on its decision to exclude inconclusive decisions from the calculation. In this paper, we also consider an FPR estimate obtained by accounting for inconclusive decisions. Thus, we consider all 42 erroneous identifications, but this time, out of the total number of decisions made, including inconclusive decisions (42 erroneous identifications out of 4536 total decisions). This calculation further decreases the FPR to 0.9%, with a 95% confidence interval upper bound of approximately 1.2%.

Although disagreements are expressed on how the FPR should be calculated, the issue of the over-representation of the 35 cases in which examiners correctly identified the individual who produced the test latent prints, but failed to associate them to the fingers from which they originated, has not been raised. Although it is clear that some type of error has occurred, it is not possible to determine a posteriori what happened. For example, these 35 cases may result from clerical errors; alternatively, they may be due to some individuals having very similar patterns across their ten fingers. We do not want to speculate on the reasons that resulted in these 35 cases; we simply consider that errors were made since the wrong conclusions were reported. However, these errors are erroneous identification to the wrong finger of the correct donor, and it is clear that, from a statistical perspective, these 35 cases need to be considered separately from erroneous identifications made to the wrong donor. The following simulations support this classification.

**IV.    Simulation Study**

*IV.a        Model of the MDPD study*

To assess the appropriateness of the error rate calculations proposed by the MDPD research team, the OSAC Friction Ridge subcommittee, and our alternative that includes inconclusive decisions, we have modelled the behaviour of the participants to the MDPD study. This model allows us to simulate the decisions and conclusions that would be obtained using different values for the rates of correct, inconclusive and erroneous conclusions when the source was present in the three sets of controls ($R_{SP}$), and when the source was not present ($R_{SA}$) (see table 2 below). Algorithms 1, 2 and 3 show the process used to simulate the Miami-Dade error rate study.

Our model is designed to keep the total number of test cases (4360 for phase 1 and 1760 for phase 2) and the number of cases where control impressions from the true sources are provided (56 out of 80 cases) fixed across all experiments. Our model is based on the four following arguments and assumptions:

(1) The MDPD report indicates that 56 (out of 80) latent prints were presented with sets of impressions from the sources that yielded them (the remaining 24 latent prints were not compared against the sets of impressions from their true sources); however, the report does not specify how many of each type of comparison were present in the phase 1 and 2 packets. To account for this lack of information, we initiate each simulation by generating a pseudo-set, $P$, of 80 latent prints where impressions from the true sources of the first 56 prints are considered to be available, and impressions from the true sources of the remaining 24 prints are considered to be absent. We then randomly sample 40 latent prints, $P_1$, from the entire set of 80 latent prints, $P$ (each of the 80 prints has an equal chance of being selected for inclusion in $P_1$). Finally, we randomly sample 20 latent prints from the remaining 40 to create set, $P_{2A}$, (each of the remaining 40 prints has an equal chance of being selected for inclusion in $P_{2A}$) and designate the final 20 latent prints as set $P_{2B}$.

(2) During phases 1 and 2 of the study, examiners were provided with a fixed set of test cases: 109 examiners were provided with 40 cases at the beginning of phase 1; 88 examiners were provided with 20 cases at the beginning of phase 2. However, only 4233 decisions were returned at the end of phase 1 (out of a total of 4360 possible decisions), and 1730 decisions were returned at the end of phase 2 (out of 1760 possible decisions). Given that some latent prints were selected to be more challenging than others and given that some examiners may have been busier than others, it is likely that the missing responses are concentrated on a small number of latent prints or a small number of examiners. However, the MDPD report does not provide information on which examiners did not finish the study and on which latent prints were favored. Assuming that all examiners are equally likely to complete the study and that all test cases are equally likely to be considered by the participants, we model the number of test cases processed by an examiner during both phases, $n_{r_i}$, by Binomial distributions (with parameters $n = 40$ cases and probability of processing a case of $p = 4233/4360$ [1] in phase 1; with parameters $n = 20$ cases and probability of processing a case of $p = 1730/1760$ in phase 2);

(3) Not all of the latent prints for which a decision was returned ended up being compared to the pre-specified sets of control prints. Some latent prints were deemed of no value and no conclusions were reported on their source (3210 latent prints were deemed of value out of 4233 decisions returned at the end of phase 1; 1342 were deemed of value out of 1730 returned at the end of phase 2). It is likely that challenging prints were more often deemed of no value; furthermore, it is also likely that some participants were more prone to deem that test cases should not be compared to control prints and that significant variability exists between the value determination of some latent prints by the participants. The MDPD report does not provide data on these elements. Assuming that all test cases for which decisions were made were equally likely to be deemed of value, and that all examiners were equally performing, we model the number of latent prints deemed of value in a given packet during phases 1 and 2 by Binomial distributions (with parameters $n = n_{r_i}$, where $n_{r_i}$ is the number of decisions returned for the $i^{th}$ packet, determined using the binomial distribution in (2) above, and probability of deeming the print of value $p = 3210/4233$ during phase 1; and, $n = n_{r_i}$ and $p = 1342/1730$ during phase 2).

---

[1] Note that it is possible to refine the model by accounting for the uncertainty on the probability of success parameter of the Binomial distribution. However, this is outside of the scope of this paper and the point estimates obtained from the data provided by MDPD are used instead.

(4) According to the design of the MDPD study, any conclusion on the source of a test case that was deemed of value for comparison must belong to one of six categories (table 1): (a) correct identification to the true donor; (b) identification to the wrong finger of the true donor; (c) identification to the wrong person; (d) inconclusive examination; (e) correct exclusion of the donors of all 3 sets of control prints; (f) incorrect exclusion of the true donor. Depending on whether control prints of the true donor are provided as part of the 3 sets of control prints, some of the categories described above cannot be used. For example, it is not possible to categorise a decision on the source of a test case as (e) when prints of the true donor are present among the provided control prints. In such case, exclusion of all donors automatically results in the decision being categorised as (f). Conversely, category (f) cannot be used when prints of the true donor are not provided.

It is likely that some of the more challenging test latent prints resulted in more inconclusive examinations than others. It is also possible that a large number of erroneous conclusions can be associated with a small number of participants. However, as mentioned above, the MDPD report does not provide data on these elements. Assuming that all latent prints are equally likely to be correctly associated to their true source and assuming that all participants are equally performing, we model the categorisation of the decision resulting from the examination of a test case for which the control prints of the true source are provided by a Multinomial distribution over the six categories (a)-(f) described above with vectors of probability parameters, $R_{SP}$, listed in table 2 below. Similarly, we model the categorisation of the decision resulting from the examination of a test case for which the control prints of the true source are absent by a Multinomial distribution over the six categories (a)-(f) described above with vectors of probability parameters, $R_{SA}$, listed in table 2 below.

Our model relies on four main assumptions:

(1) All 80 latent prints are equally likely to be selected for inclusion in phase 1 and phase 2 packets.

(2) All examiners are equally performing;

(3) All latent prints are equally likely to be deemed of value;

(4) All latent prints are equally likely to result in an error (false positive or false negative).

Based on the MDPD report, there is no indication that assumption (1) is unreasonable. All research to date on the performance of latent print examiners shows that assumption (2) is unlikely to be correct. Furthermore, by the design of the MDPD study, assumptions (3) and (4) are equally incorrect since they have purposefully selected test prints with various levels of quality. However, we are interested in estimating the average FPR for a population of examiners and for a range of latent print quality. Thus, we believe that these assumptions are reasonable enough to enable us to estimate the average FPR. The appropriateness of these assumptions is tested below by using the model to repeatedly simulate the MDPD study and comparing the results of the simulations with the data observed by the MDPD project team. Because the results presented in table 3 are similar to those presented in table 1, we can conclude that, although incorrect, the assumptions are fit-for-purpose in that they allow the model to replicate the MDPD study.

**Algorithm 1:** Simulation to reproduce phase 1 of the Miami-Dade study

**Define:** $R_{SP}$, a set of point estimates for the classification rates when control impressions from the true source are present in the comparison process; $R_{SA}$, a set of point estimates for the classification rates when control impressions from the true source are not provided (absent) in the comparison process; $P$, a set of 80 latent prints (to correspond to the MDPD study (p. 28) we consider, without loss of generality, that control impressions from the true source are provided for prints 1 through 56, and are not provided for prints 57 through 80);

**for** Phase 1 **do**
    Sample 40 latent prints, $P_1$, from $P$;
    **for** $i \in 1 : 109$ packets **do**
        Sample $n_{r_i} \sim \text{Binomial}\left(40, \frac{4233}{4360}\right)$, the number of decisions that will be returned in packet $i$;
        Sample $n_{v_i} \sim \text{Binomial}\left(n_{r_i}, \frac{3210}{4233}\right)$, the number of decisions that are determined to be of value in packet $i$;
        Sample $n_{v_i}$ latent prints, $P_{n_{v_i}}$, from $P_1$ to determine which prints are considered in packet $i$;
        Sample $d_{SP_i} \sim \text{Multinomial}\left(1, \sum_{P_1} I\{P_{n_{v_i}} \in P_{SP_i}\}, R_{SP}\right)$, the decisions made for $P_{SP_i}$, the set of latent prints in packet $i$ whose source is provided;
        Sample $d_{SA_i} \sim \text{Multinomial}\left(1, \sum_{P_1} I\{P_{n_{v_i}} \in P_{SA_i}\}, R_{SA}\right)$, the decisions made for $P_{SA_i}$, the set of latent prints in packet $i$ whose source is not provided;
    **end**
    Calculate $n_{SP}^{(1)} = \sum_{i=1}^{109} d_{SP_i}$, the total count for each decision classification when the source is present;
    Calculate $n_{SA}^{(1)} = \sum_{i=1}^{109} d_{SA_i}$, the total count for each decision classification when the source is not present;
    **Return:** $n_{SP}^{(1)}$ and $n_{SA}^{(1)}$.
**end**

---

**Algorithm 2:** Simulation to reproduce phase 2 of the Miami-Dade study

**Define:** $R_{SP}$, a set of point estimates for the classification rates when control impressions from the true source are present in the comparison process; $R_{SA}$, a set of point estimates for the classification rates when control impressions from the true source are not present (absent) in the comparison process; $P_2 := P \backslash P_1$;

**for** Phase 2 **do**
    Sample 20 latent prints, $P_{2A}$, from $P_2$;
    Define $P_{2B}$, the remaining set of 20 latent prints;
    **for** group $k \in \{A, B\}$ **do**
        **for** $i \in 1 : 44$ packets **do**
            Sample $n_{r_{ik}} \sim \text{Binomial}\left(20, \frac{1730}{1760}\right)$, the number of decisions that will be returned in packet $i$ for group $k$;
            Sample $n_{v_{ik}} \sim \text{Binomial}\left(n_{r_{ik}}, \frac{1342}{1730}\right)$, the number of decisions that are determined to be of value in packet $i$ for group $k$;
            Sample $n_{v_{ik}}$ latent prints, $P_{n_{v_{ik}}}$, from $P_{2k}$ to determine which prints are considered in packet $i$ for group $k$;
            Sample $d_{SP_{ik}} \sim \text{Multinomial}\left(1, \sum_{P_{2k}} I\{P_{n_{v_{ik}}} \in P_{SP_{ik}}\}, R_{SP}\right)$, the decisions made for $P_{SP_{ik}}$, the set of latent prints in packet $i$ for group $k$ whose source is provided;
            Sample $d_{SA_{ik}} \sim \text{Multinomial}\left(1, \sum_{p_1} I\{P_{n_{v_{ik}}} \in P_{SA_{ik}}\}, R_{SA}\right)$, the decisions made for $p_{SA_{ik}}$, the set of latent prints in packet $i$ for group $k$ whose source is not provided;
        **end**
    **end**
    Calculate $n_{SP}^{(2)} = \sum_{k \in \{A,B\}} \sum_{i=1}^{44} d_{SP_{ik}}$, the total count for each decision classification when the source is present;
    Calculate $n_{SA}^{(2)} = \sum_{k \in \{A,B\}} \sum_{i=1}^{44} d_{SA_{ik}}$, the total count for each decision classification when the source is not present;
    **Return:** $n_{SP}^{(2)}$ and $n_{SA}^{(2)}$.
**end**

**Algorithm 3:** Simulation to reproduce Miami-Dade study

**Define:** $R_{SP}$, a set of point estimates for the classification rates when control impressions from the true source are present in the comparison process; $R_{SA}$, a set of point estimates for the classification rates when when control impressions from the true source are not present (absent) in the comparison process; $p$, a set of 80 latent prints; $N$, the number of simulations (typically $N \geq 1000$);

**for** $i \in 1{:}N$ **do**
    Simulate phase 1 according to algorithm 1, with $R_{SP}$, $R_{SA}$, $P$ and $N$ as defined above;
    Simulate phase 2 according to algorithm 2, with $R_{SP}$, $R_{SA}$, $P$ and $N$ as defined above;
**end**

Calculate $\bar{n}_{SP} = \frac{1}{N} \sum_{i=1}^{2} \sum_{j=1}^{N} n_{SP_j}^{(i)}$, the mean count of each decision classification when the source is present;

Calculate $\bar{n}_{SA} = \frac{1}{N} \sum_{i=1}^{2} \sum_{j=1}^{N} n_{SA_j}^{(i)}$, the mean count of each decision classification when the source is not present;

**Return:** $\bar{n}_{SP}$ and $\bar{n}_{SA}$.

### IV.b  Simulations

We report below the results from a total of seven experiments involving simulations with different rate vectors, $R_{SP}$ and $R_{SA}$, for the categorisation of the conclusions regarding the sources of the test cases. The first simulation uses point estimates of the rates calculated from the observed frequencies reported by MDPD (see table 1). This experiment is designed to verify that our model is able to reproduce the observations made by the MDPD research team, despite the simplifications and assumptions discussed in the previous section.

The remaining six experiments are associated with the three error rates proposed by the MDPD, the OSAC FRS, and our alternative to the OSAC FRS estimate that includes inconclusive decisions. We do not consider the 4.2% FPR proposed by MDPD, or any of the upper bounds on the MDPD FRP estimates proposed by PCAST [4, p. 96] for reasons that will become clear later. We also do not formally test the estimates proposed by Ulery et al. [1], as they are close to the estimates proposed by OSAC FRS, and considering both sets of results would be redundant. We note that, aside from the first set of rates (based on the observed frequencies from the MDPD study), a single FPR is considered for all simulated comparisons, regardless of whether or not (a) the true source was present during the comparison process, or (b) the examiner identified the correct person, but the wrong finger. In our simulations, the chosen FPR corresponds to the error rate estimates proposed by MDPD, the OSAC FRS, and our alternative to the OSAC FRS estimate. The use of a single FPR relies on the assumption that every latent print has the same chance to result in an erroneous identification[2].

For each of the three proposed error rates, we conduct two experiments: in the first experiment, we consider the proposed FPRs described in section III, and two inconclusive rates based on the frequencies observed in the MDPD study (446/3177 and 403/1359 in table 1); in the second experiment, we consider the same FPRs, and a common inconclusive rate based on the pooled frequencies observed in the MDPD study (849/4536 in table 1).

Finally, the same false negative rate (FNR) is used throughout all six experiments. We used the one reported by MDPD [3, table 5]. The true positive rate (TPR) is calculated by subtracting the sum of all other "source present" rates from one (e.g., for the Miami-Dade Proposed rates, the TPR is calculated by $1 - (0.030 + 0.075 + 0.140) = 0.755$). Likewise, the true negative rate (TNR) is calculated by subtracting the sum of all other "source not present" rates from one (e.g., for the Miami-Dade proposed rates, the TNR is calculated by $1 - (0.030 + 0.30) = 0.0667$). The various rates associated with our experiments are reported in table 2.

In each experiment, we simulate 1,000 iterations of the MDPD study to estimate expected counts for each category of decision as well as a 95% credible interval (table 3).

---

[2] This assumption is not reasonable: lower quality latent prints have been shown to result in higher error rates [7]. However, the key question raised by the MDPD, as we will discuss below, is whether the rate of erroneous identifications is the same when control impressions are provided as when they are not.

| Proposed Rates | True Source Present (Yes: $R_{SP}$/No: $R_{SA}$) | FPR | | FNR | TPR | TNR | Inc. |
|---|---|---|---|---|---|---|---|
| | | Correct person, wrong finger | Incorrect person | | | | |
| **Observed frequencies** | Yes | 0.011 | 0.001 | 0.074 | 0.774 | N/A[4] | 0.140 |
| | No | N/A[1] | 0.002 | N/A[2] | N/A[3] | 0.698 | 0.300 |
| **Miami-Dade** | Yes | 0.030 | | 0.075 | 0.755 | N/A | 0.140 |
| | No | | | N/A | N/A | 0.667 | 0.300 |
| **Miami-Dade (common inc.)** | Yes | | | 0.075 | 0.708 | N/A | 0.187 |
| | No | | | N/A | N/A | 0.783 | |
| **OSAC FRS** | Yes | 0.011 | | 0.075 | 0.774 | N/A | 0.140 |
| | No | | | N/A | N/A | 0.689 | 0.300 |
| **OSAC FRS (common inc.)** | Yes | | | 0.075 | 0.727 | N/A | 0.187 |
| | No | | | N/A | N/A | 0.802 | |
| **OSAC FRS alternative** | Yes | 0.009 | | 0.075 | 0.776 | N/A | 0.140 |
| | No | | | N/A | N/A | 0.691 | 0.300 |
| **OSAC FRS alternative (common inc.)** | Yes | | | 0.075 | 0.729 | N/A | 0.187 |
| | No | | | N/A | N/A | 0.804 | |

**Table 2:** Rates considered in simulation experiments. [1] It is not possible to identify the wrong finger of the correct person when control prints from the true sources are not present. [2] It is not possible to erroneously exclude the correct source when control prints from the true sources are not present. [3] It is not possible to correctly associate the test latent prints with control prints from their true sources when those are not present. [4] The design of the MDPD study made it impossible to have a rate of correct exclusion of a source when the true sources were provided in the test packets.

### IV.c    Results of the experiment using point estimates from MDPD report

To assess the reasonableness of the assumptions underlying our model, we begin by defining the set $R \coloneqq \{R_{SP}, R_{SA}\}$ of classification rates according to the rates specified in the *Observed Frequencies* row of table 2, such that $R_{SP} \coloneqq \{0.011, 0.001, 0.074, 0.774, 0, 0.140\}$ and $R_{SA} \coloneqq \{0, 0.002, 0, 0, 0.698, 0.300\}$. It is not possible to observe all of the categories presented in table 2 (e.g., it is not possible to have a correct exclusion when the source is present in the comparison process, a correct identification when the source is not present in the comparison process, an erroneous exclusion when the source is not present in the comparison process), and so the unobservable categories are indicated in the tables throughout the paper by N/A, and are assigned a zero probability in the set $R$.

When comparing the range of frequencies resulting from our experiment (table 3) to the actual frequencies observed by MDPD (table 1), we can see that our algorithm produces a reasonable model of the MDPD study, despite its assumptions and approximations. Thus, we proceed by using it to investigate the reasonableness of the proposed error rates.

|  |  | Source present | Source not present | Totals |
|---|---|---|---|---|
| # of Latent Prints | | 56 | 24 | 80 |
| # of Decisions | | 3188.48 [2983, 3369] | 1362.71 [1169, 1563] | 4551.19 [4479, 4612] |
| Correct IDs | | 2466.29 [2302, 2620] | N/A | 2466.29 [2302, 2620] |
| Erroneous IDs | Correct person, wrong finger | 35.38 [24, 47] | N/A | 35.38 [24, 47] |
| | Incorrect person | 4.06 [0, 7] | 2.98 [0, 6] | 7.04 [1, 11] |
| Inconclusive examinations | | 447.41 [396, 490] | 404.72 [335, 468] | 852.12 [800, 918] |
| Correct Exclusions | | N/A | 955.01 [806, 1097] | 955.01 [806, 1097] |
| Erroneous Exclusions | | 235.34 [202, 266] | N/A | 235.34 [202, 266] |

**Table 3:** Reproduced Miami-Dade results using simulations and plug-in estimates for the rates of the various conclusions calculated using the values in table 1. The numbers in [ ] represent the lower/upper bounds of maximum density 95% credible intervals.

*IV.d   Results of the experiments using the MDPD proposed FPR of 3.0%*

To test the MDPD proposed FPR of 3.0%, rate vectors $R_{SP}, R_{SA}$ are defined according to the rates specified in the *Miami-Dade* and *Miami-Dade (common inc.)* rows of table 2. In the first experiment, we use the explicit rates proposed by Pacheco et al. [3]. In the second experiment, we consider a common inconclusive rate. This enables us to test the consistency of the inconclusive rate depending on whether control prints from the true sources are provided.

When using the 3.0% FPR suggested by MDPD, the average number of erroneous identifications that are produced by the model (tables 4 and 5) is remarkably larger than the number of actual erroneous identifications observed by the MDPD team (table 1): table 1 reports a total of 42 erroneous identifications; table 4 indicates that the number of erroneous identifications should have been between 109 and 157 with 0.95 probability, if the FPR was truly 3.0%; and table 5 indicates that the number of erroneous identifications should have been between 115 and 159 with 0.95 probability, if the FPR was truly 3.0%.

Furthermore, we note that reporting a single inconclusive rate is not a reasonable interpretation of the MDPD results. Although the total number of inconclusive decisions is consistent between

|  |  | Source present | Source not present | Totals |
|---|---|---|---|---|
| # of Latent Prints |  | 56 | 24 | 80 |
| # of Decisions |  | 3187.07 [3003, 3392] | 1364.26 [1163, 1533] | 4551.33 [4481, 4612] |
| Correct IDs |  | 2405 [2252, 2554] | N/A | 2405.07 [2252, 2554] |
| Erroneous IDs | Correct person, wrong finger | N/A | N/A | N/A |
|  | Incorrect person | 95.85 [76, 116] | 40.74 [27, 54] | 136.59 [109, 157] |
| Inconclusive examinations |  | 447.54 [402, 494] | 403.73 [339, 468] | 851.28 [785, 902] |
| Correct Exclusions |  | N/A | 919.79 [791, 1051] | 919.79 [791, 1051] |
| Erroneous Exclusions |  | 238.61 [209, 270] | N/A | 238.61 [209, 270] |

**Table 4:** Average number of conclusions in each category using Miami-Dade FPR (3.0%) and FNR (7.5%) – with different rates of inconclusive examinations. The numbers in [ ] represent the lower/upper bounds of maximum density 95% credible intervals from our 1,000 simulations.

tables 1, 4 and 5, the number of inconclusive decisions when the source is present in the comparison process is larger in our experiment, and the number of inconclusive decisions when the source is not present in the comparison process is much smaller in our experiment.

The implications of these results are two-fold: first, it is clear that 3.0% is an over-estimation of the true FPR of the participants of the MDPD study; second, examiners behave differently when impressions from the true sources are present in the comparison process. When control impressions are present, examiners make an inconclusive decision approximately 14% of the time, compared to approximately 30% of the time when control impressions are not present.

|  |  | Source present | Source not present | Totals |
|---|---|---|---|---|
| # of Latent Prints |  | 56 | 24 | 80 |
| # of Decisions |  | 3188.44 [2986, 3369] | 1364.26 [1165, 1549] | 4552.70 [4490, 4616] |
| Correct IDs |  | 2256.75 [2107, 2401] | N/A | 2256.75 [2107, 2401] |
| Erroneous IDs | Correct person, wrong finger | N/A | N/A | N/A |
|  | Incorrect person | 95.50 [74, 114] | 41.02 [28, 55] | 136.53 [115, 159] |
| Inconclusive examinations |  | 597.33 [539, 648] | 255.81 [213, 300] | 853.13 [802, 904] |
| Correct Exclusions |  | N/A | 1067.43 [905, 1210] | 1067.43 [905, 1210] |
| Erroneous Exclusions |  | 238.86 [205, 269] | N/A | 238.86 [205, 269] |

**Table 5:** Average number of conclusions in each category using Miami-Dade FPR (3.0%) and FNR (7.5%) – with common rate of inconclusive examinations. The numbers in [ ] represent the lower/upper bounds of maximum density 95% credible intervals from our 1,000 simulations.

Since the FPR estimate proposed by MDPD is already significantly larger than the true FPR of the participants in their study (the lower bound of our 95% credible interval on the number of erroneous identifications that should have been observed is much larger than the 42 erroneous identifications observed by MDPD), we did not repeat the experiment using the upper bound of the confidence interval for the FPR (5.4%) proposed by PCAST [4, p. 95]. Indeed, the resulting observations would be superfluous.

*IV.e     Results of the experiments using the OSAC FRS proposed FPR of 1.1%*

To test the FPR of 1.1% proposed by the OSAC FRS, we define *R* according to the rates specified in the *OSAC FRS* and *OSAC FRS (common inc.)* rows of table 2. While we define the FPR according to the OSAC FRS's proposed error rate, we use the same FNR and inconclusive rates that are obtained from the MDPD study.

The results of these experiments show that the average number of erroneous identifications, obtained through our simulations (52 expected erroneous identifications in tables 5 and 6) is slightly higher than the 42 observed by MDPD (table 1). Since 42 is within the 95% credible intervals for both experiments ([39, 65] in table 6 and [37, 65] in table 7), it may appear, at first, that an FPR of 1.1% for this study is not unreasonable. However, our results show that, while considering an FPR of 1.1% is reasonable when control prints from the true sources are presented with the test cases, this FPR is an overestimate of the true FPR when control prints from the true sources are not present: indeed, the number of erroneous identifications reported in table 1 (3 erroneous identifications) is largely outside of the credible intervals resulting from our simulations ([7,23] in table 6 and [8, 23] in table 7). Thus, the FPR of 1.1% proposed by the OSAC FRS is not necessarily a fair estimate of the true FPR of the participants in the MDPD study and our results suggest again that a single FPR may not be appropriate to interpret the MDPD observations.

Finally, we confirm that considering a single inconclusive rate for the MDPD study is not reasonable: the average number of inconclusive decisions when control prints from the true sources are present (596 in table 7) is greater than the corresponding number in table 1 (446), while the number of inconclusive decisions when control prints from the true sources are not present (256 in table 7) is smaller than the corresponding number in table 1 (403).

|  |  | Source present | Source not present | Totals |
|---|---|---|---|---|
| # of Latent Prints | | 56 | 24 | 80 |
| # of Decisions | | 3185.49 [3000, 3405] | 1365.77 [1157, 1555] | 4551.26 [4486, 4617] |
| Correct IDs | | 2462.62 [2298, 2633] | N/A | 2462.62 [2298, 2633] |
| Erroneous IDs | Correct person, wrong finger | N/A | N/A | N/A |
| | Incorrect person | 36.35 [24, 47] | 15.49 [7, 23] | 51.84 [39, 65] |
| Inconclusive examinations | | 447.78 [402, 498] | 405.75 [342, 473] | 853.52 [797, 912] |
| Correct Exclusions | | N/A | 944.53 [808, 1091] | 944.53 [808, 1091] |
| Erroneous Exclusions | | 238.74 [206, 272] | N/A | 238.74 [206, 272] |

**Table 6:** Average number of conclusions in each category using OSAC FRS (1.1%) and FNR (7.5%) – with different rates of inconclusive examinations. The numbers in [ ] represent the lower/upper bounds of maximum density 95% credible intervals from our 1,000 simulations.

|  |  | Source present | Source not present | Totals |
|---|---|---|---|---|
| # of Latent Prints | | 56 | 24 | 80 |
| # of Decisions | | 3184.69 [2986, 3373] | 1368.15 [1186, 1562] | 4552.85 [4484, 4616] |
| Correct IDs | | 2313.27 [2167, 2467] | N/A | 2313.27 [2167, 2467] |
| Erroneous IDs | Correct person, wrong finger | N/A | N/A | N/A |
| | Incorrect person | 36.39 [25, 47] | 15.66 [8, 23] | 52.05 [37, 65] |
| Inconclusive examinations | | 596.21 [540, 652] | 256.50 [206, 297] | 852.71 [803, 904] |
| Correct Exclusions | | N/A | 1095.99 [943, 1256] | 1095.99 [943, 1256] |
| Erroneous Exclusions | | 238.83 [208, 271] | N/A | 238.83 [208, 271] |

**Table 7:** Average number of conclusions in each category using OSAC FRS (1.1%) and FNR (7.5%) – with common rate of inconclusive examinations. The numbers in [ ] represent the lower/upper bounds of maximum density 95% credible intervals from our 1,000 simulations.

### IV.f     Results of the experiments using the alternative FPR of 0.9%

To test the FPR of 0.9%, an alternative to the OSAC FRS estimate which accounts for inconclusive decisions, we define *R* according to the rates specified in the *OSAC FRS alternative* and *OSAC FRS alternative (common inc.)* rows of table 2. We use the same FNR and inconclusive rates as in the previous experiments.

In tables 8 and 9, we observe that the average number of erroneous identifications obtained from the simulations is very similar to the 42 observed by MDPD. Nevertheless, and contrary to the experiments performed using the 1.1% FPR proposed by OSAC FRS, we observe that, in this case, our simulations result in an average number of erroneous identification when control prints form the true sources are present that is lower than the 39 observed by MDPD. Additionally, the average number of erroneous identifications when control prints from the true sources are absent is higher than the 3 observed by MDPD. Once again, our results suggest that a single FPR shared by the two scenarios may not be appropriate in this study and that examiners may behave differently when control prints from the true sources are presented together with the test latent prints.

Lastly, as noted in the two previous experiments, the results of this experiment lead to the conclusion that considering a single inconclusive rate is not reasonable for this study.

|  |  | Source present | Source not present | Totals |
|---|---|---|---|---|
| # of Latent Prints | | 56 | 24 | 80 |
| # of Decisions | | 3188.56 [3004, 3378] | 1363.59 [1184, 1552] | 4552.15 [4486, 4618] |
| Correct IDs | | 2472.12 [2327, 2627] | N/A | 2472.12 [2327, 2627] |
| Erroneous IDs | Correct person, wrong finger | N/A | N/A | N/A |
| | Incorrect person | 29.49 [19, 40] | 12.62 [5, 19] | 42.11 [29, 54] |
| Inconclusive examinations | | 447.61 [399, 492] | 403.35 [337, 461] | 850.96 [789, 908] |
| Correct Exclusions | | N/A | 947.62 [815, 1088] | 947.62 [815, 1088] |
| Erroneous Exclusions | | 239.35 [204, 269] | N/A | 239.35 [204, 269] |

**Table 8:** Average number of conclusions in each category using the FPR that accounts for inconclusive decisions (0.9%) and FNR (7.5%) – with different rates of inconclusive examinations. The numbers in [ ] represent the lower/upper bounds of maximum density 95% credible intervals from our 1,000 simulations.

|  |  | Source present | Source not present | Totals |
|---|---|---|---|---|
| # of Latent Prints | | 56 | 24 | 80 |
| # of Decisions | | 3188.87 [2984, 3373] | 1363.74 [1169, 1552] | 4552.62 [4492, 4620] |
| Correct IDs | | 2322.80 [2174, 2479] | N/A | 2322.80 [2174, 2479] |
| Erroneous IDs | Correct person, wrong finger | N/A | N/A | N/A |
| | Incorrect person | 29.09 [18, 38] | 12.58 [5, 19] | 41.67 [28, 54] |
| Inconclusive examinations | | 596.95 [542, 656] | 255.11 [209, 299] | 852.06 [798, 901] |
| Correct Exclusions | | N/A | 1096.06 [936, 1256] | 1096.06 [936, 1256] |
| Erroneous Exclusions | | 240.03 [206, 269] | N/A | 240.03 [206, 269] |

**Table 9:** Average number of conclusions in each category using the FPR that accounts for inconclusive decisions (0.9%) and FNR (7.5%) – with common rate of inconclusive examinations. The numbers in [ ] represent the lower/upper bounds of maximum density 95% credible intervals from our 1,000 simulations.

## V. Discussion and Conclusions

The interpretation of the data resulting from a study conducted by the Miami Dade Police Department (MDPD) and designed to estimate error rates in fingerprint examination [3] has generated some controversy. The core of the controversy resides in that the false-positive error rate (FPR) of 3.0% proposed by the MDPD research team results from the ratio of two numbers observed under two different scenarios: (a) the total number of erroneous identifications made during the entire study; (b) the number of decisions rendered when control prints of the true sources of the test cases were not provided. In other words, the research team did not fully account for the decisions rendered when control prints of the true sources were provided.

As a result of the MDPD study, false-positive error rate (FPR) estimates have been proposed by the MDPD, and the OSAC Friction Ridge subcommittee [6]. Furthemore, an additional calculation based on the one from the OSAC Friction Ridge subcommittee that includes inconclusive decisions is considered. In this paper, we model the behaviour of the participants to the MDPD study in order to simulate the number of erroneous identifications that should be observed if the true FPR was close to any of the proposed estimates. While our model relies on necessary simplifications and assumptions, the results of our first experiment show that the model can generate data that is very similar to the observations made by MDPD (table 3), and that it can be considered to be adequate. Overall, our different experiments show that none of the proposed FPRs are good estimates of the true FPR of the study's participants:

(1) the 3.0% FPR proposed by MDPD largely overestimates the true FPR (tables 4 and 5);

(2) the 1.1% FPR proposed by the OSAC FRS [6] seems larger than the FPR of the participants examining test cases that were not associated with control prints from the true sources of the latent prints (tables 6 and 7);

(3) the 0.9% FPR that accounts for inconclusive decisions appears to be marginally smaller than the FPR of the participants examining test cases that were associated with control prints from their true sources, but marginally larger than the FPR of the participants examining test cases that were not associated with control prints from the true sources (tables 8 and 9).

Although the OSAC FRS [6] and the Canadian case study [5] were correct to point out the mistake made by the MDPD and PCAST when calculating their estimates of the FPR, their analyses of the original MDPD error rate estimate and their attempts to resolve the calculation for the FPR are still incorrect. The FPR calculations proposed by the OSAC FRS [6] and the alternative that accounts for inconclusive decisions certainly provide better estimates; yet, our simulations show that neither of them is satisfactory.

The main difficulty with the interpretation of the MDPD data is that participants to their study seem to have two distinctly different levels of performance depending on whether control prints from the true sources of the test latent prints are provided or not. This observation stems from the wildly different estimates of the rates at which participants deem that examinations are inconclusive (14% when the control prints from the true source are provided vs. 30% when they are not) and at which erroneous identifications occur (~1.2% when the control prints from the true source are provided vs. ~0.1% when they are not) (see tables 1 and 2). From this latter observation, and from the results of our experiments, it appears that reporting a single FPR and a single rate of inconclusive examinations based on the data acquired during the MDPD study is inappropriate.

To properly interpret the data acquired by the MDPD and reported in table 1, we propose two alternative solutions. Both solutions consider that the participants to the study have two distinct true FPRs:

(1) First, we can consider that examiners do make more false identifications when they are unknowingly provided with control impressions from the true sources than when they are not. This solution is counterintuitive but explains the data. In this case, a Bayesian analysis of the false identification rates (using a flat Beta prior distribution on the FPR parameter and a binomial likelihood) assigns an upper bound for the 95% credible interval of the FPR of 1.7% when impressions of the true source are provided, and of 0.6% when impressions of the true source are not provided. The maximum a posteriori (MAP) estimates for these two intervals are 1.2% and 0.2%, respectively.

(2) Second, we may consider that there are two types of false identifications, consisting in (a) false identifications made to the correct person, but to the incorrect finger; and (b) false identifications made to the incorrect person regardless of whether the examiner is presented with impressions from the true sources[3]. This distinction between false identifications parallels the argument made by Koehler, that "not all false positive errors are equal" [8, p. 1412]. A Bayesian analysis of these two scenarios results in an upper bound for the 95% credible interval of 1.1% chance to erroneously identify the wrong finger of the correct source, and of 0.3% chance to erroneously identify the wrong person. The MAP estimates for these two intervals are 0.7% and 0.2%, respectively.

Both of our solutions result in FPR estimates that are much more comparable to the ones proposed by Ulery et al. [1]. We note that the worst-case estimate resulting from our interpretation of the MDPD data (1.7%) is still much lower than the PCAST 5.4% upper bound on the MDPD estimate of 4.2%.

Our analysis of the MDPD data raises questions related to the dependability of examiners' conclusions. The design of the MDPD study certainly results in a more challenging analysis of the data and to some controversy regarding the estimation of the FPR; nonetheless, it allows for highlighting an important issue that did not appear in the Ulery et al. [1] study; namely, the large number of erroneous identifications made to the wrong finger of the correct person. We are deeply concerned with the very large number of cases (35 out of 3177 decisions) where the test latent prints were associated with the wrong finger of the correct source, in particular when this number is compared to the number of erroneous identifications to the wrong source (7 out of 4536 decisions). As we mentioned previously, we do not want to speculate on the reason(s) behind such a large discrepancy. We also realise that the experiment did not account for the verification stage, which would, hopefully, catch clerical errors. However, the MDPD data shows that the number of this type of errors far outweigh the number of erroneous identifications to the incorrect person. If these errors are true clerical errors, we believe that they could easily be avoided in the first place. If they are the manifestation of a more complex issue, adequate training and procedures should be designed to remedy the problem. Similarly, we are concerned with the large discrepancy between the two distinct rates of inconclusive examinations observed during the MDPD study. This indicates that the exclusion process is different than the identification process and that it is not necessarily well understood and implemented in practice. Overall, we believe that it is critical to investigate and address the reasons behind these different behaviours of the fingerprint examiners and limit their impacts on the fingerprint examination process.

---

[3] It is trivial enough to show that there is no statistically significant differences between the two rates of erroneous identification to the incorrect person based on the estimates of 4/3177 and 3/1359.

**References**

[1] Ulery, Bradford T. R., Austin Hicklin, JoAnn Buscaglia, and Maria Antonia Roberts. 2011. "Accuracy and reliability of forensic latent fingerprint decisions." Proceedings of the National Academy of Sciences 108(19): 7733-7738.

[2] Langenburg, Glenn, 2012. "A Critical Analysis and Study of the ACE-V Process", University of Lausanne, Switzerland, PhD Thesis, 362p. https://www.unil.ch/files/live/sites/esc/files/shared/Langenburg_Thesis_Critical_Analysis_of_ACE-V_2012.pdf

[3] Pacheco, Igor. Brian Cerchiai and Stephanie Stoiloff. 2014. "Miami-Dade Research Study for the Reliability of the ACE-V Process: Accuracy & Precision in Latent Fingerprint Examinations." https://www.ncjrs.gov/App/Publications/abstract.aspx?ID=270637.

[4] President's Council of Advisors on Science and Technology (PCAST). 2016. *REPORT TO THE PRESIDENT Forensic Science in Criminal Courts: Ensuring Scientific Validity of Feature-Comparison Methods.* Washington, D.C.: Executive Office of the President's Council of Advisors on Science and Technology.

[5] Wilkinson, Della, David Richard and Daniel Hockey. 2018. "Expert Fingerprint Testimony Post-PCAST - A Canadian Case Study." Journal of Forensic Identification. 68(3): 299-331.

[6] Organization of Scientific Area Committees (OSAC) Friction Ridge Subcomittee. 2016. "Response to the President's Council of Advisors on Science and Technology's (PCAST) request for additional references." https://www.nist.gov/sites/default/files/documents/2016/12/16/osac_friction_ridge_subcommittees_response_to_the_presidents_council_of_advisors_on_science_and_technologys_pcast_request_for_additional_references_-_submitted_december_14_2016.pdf.

[7] Ulery, Bradford T., R. Austin Hicklin, Maria Antonia Roberts, and JoAnn Buscaglia. 2014. "Measuring What Latent Fingerprint Examiners Consider Sufficient Information for Individualization Determinations." PLS ONE 9(11): e110179. https://doi.org/10.1371/journal.pone.0118172.

[8] Koehler, J.J. 2017. "Forensics or Fauxrensics? Ascertaining Accuracy in the Forensic Sciences." Arizona State Law Journal 49: 1369-1416.